# Growth of Transition Metal Dichalcogenides by Solvent Evaporation Technique

Dmitriy A. Chareev[1,2,3*], Polina V. Evstigneeva[4], Dibya Phuyal[5], Gabriel J. Man[5], Håkan Rensmo[5], Alexander N. Vasiliev[3,6,7], and Mahmoud Abdel-Hafiez[5,8*]

[1]Institute of Experimental Mineralogy, 142432, Chernogolovka, Russia

[2]Kazan Federal University, Kazan, 420008, Russia

[3]Ural Federal University, Yekaterinburg 620002, Russia

[4]Institute of Geology of Ore Deposits, Petrography, Mineralogy and Geochemistry, Moscow, 119017, Russia

[5]Department of Physics and Astronomy, Uppsala University, Uppsala, SE-75120, Sweden

[6]Lomonosov Moscow State University, Moscow 119991, Russia

[7]National Research South Ural State University, Chelyabinsk 454080, Russia

[8]Kirchhoff Institute of Physics, Heidelberg University, D-69120 Heidelberg, Germany

ABSTRACT: Due to their physical properties and potential applications in energy conversion and storage, transition metal dichalcogenides (TMDs) have garnered substantial interest in recent years. Amongst this class of materials, TMDs based on molybdenum, tungsten, sulfur and selenium are particularly attractive due to their semiconducting properties and the availability of bottom-up synthesis techniques. Here we report a method which yields high quality crystals of transition metal diselenide and ditelluride compounds ($PtTe_2$, $PdTe_2$, $NiTe_2$, $TaTe_2$, $TiTe_2$, $RuTe_2$, $PtSe_2$, $PdSe_2$, $NbSe_2$, $TiSe_2$, $VSe_2$, $ReSe_2$) from their solid solutions, via vapor deposition from a metal-saturated chalcogen melt. Additionally, we show the synthesis of rare-earth metal poly-chalcogenides and $NbS_2$ crystals using the aforementioned process. Most of the obtained crystals have a layered $CdI_2$ structure. We have investigated the physical properties of selected crystals and compared them to state-of-the-art findings reported in the literature. Remarkably, the charge density wave transition in 1T-$TiSe_2$ and 2H-$NbSe_2$ crystals is well-defined at $T_{CDW}$ ~ 200 K and ~ 33 K, respectively. Angle-resolved photoelectron spectroscopy and electron diffraction are used to directly access the electronic and crystal structures of $PtTe_2$ single crystals, and yield state-of-the-art measurements.

## INTRODUCTION

Transition metal dichalcogenides (TMDs), due to the interplay between electronic correlations and electronic orderings provide an excellent platform to investigate many-body physics. Their structural anisotropy results in highly anisotropic physical properties and induces a quasi-one-dimensional behavior. The low-dimensional superconducting systems, in particular, transition metal dichalcogenides, $MX_2$, with $CdI_2$ structure (M - transition metals IV-VII of the Ti, Nb, Mo, Ta, W, Pd, Pt and X = Se, S, Te), shown a coexistence of superconductivity and competing physical phenomena such as magnetic, charge order, and charge density waves, (CDW). T This coexistence has been of interest for the condensed matter community for a long time [1,2]. The crystals of these compounds consist of layers, each of which is a sandwich of two layers of halogen atoms X with a layer of metal atoms M between them. The bond between metal atoms and halogens in a sandwich is strong (predominantly covalent), and the M and X atoms in a sandwich form a two-dimensional hexagonal lattice. Between the layers of $MX_2$ are connected in the crystal by weak van der Waals forces. The weak van der Waals coupling between the layers in dichalcogenides allows impurity atoms or molecules to be introduced into the space between these layers. This leads to an increase in the distance between the conducting layers and a significant decrease in the overlap of their electron wave functions, as a result of which the conductivity anisotropy can increase by several orders of magnitude [3]. Taking into account the superconductivity discovered in 2008 in chalcogenides and iron pnictides, it is interesting that the introduction of impurity tends to a certain increase in the critical temperature $T_c$ and a significant increase in the anisotropy of the upper critical field [4]. Recently, several new works have been reported in which the structural, electronic, and superconducting properties of the $TiSe_2$ [5], $WTe_2$ [6], $TaS_2$ and $TaSe_2$ [7,8] , $ReS_2$ [9], $SnSe_2$ [10] compounds were investigated high pressures up to 50 GPa.

The analysis of phase diagrams of the respective binary systems [11] shows that most of layered dichalcogenides melt incongruently (the exceptions are $AuTe_2$, $SnS_2$, $SnSe_2$). Therefore, it is hard to synthesize crystals of those compounds using melt growth techniques. Gas transport, hydrothermal and flux techniques can be used to synthesize such crystals. The main advantage of incongruent techniques is the possibility to obtain crystals that are not in equilibrium with their corresponding melts. One of the main basics on gas transport method is that the reversible chemical reaction of the initial solid material in the form of powder (charge) with a transport reagent. This leads to the formation of only gaseous products. This gas transported to another part of the reaction system with other physicochemical conditions, where a crystal of initial material grows. For the growth of dichalcogenide crystals, halogens or their chemical compounds are used as transport reagents [12, 13].

Flux crystal growth technique [14,15] usually consists in gradual cooling of a multicomponent flux. As the temperature decreases, the solubility of the components reduces and leads to crystallization of certain substances. Moreover, oversaturation can be created by a temperature gradient. For example, dichalcogenide crystals can be obtained in eutectic melts of alkali metal salts in a steady-state temperature gradient. The reaction is conducted in a quartz glass ampoule, which is placed in the furnace in a temperature gradient. The temperature of the hot end was 850-400°C, the temperature of the cold end was approximately 100-70 degrees lower. The temperature of crystal growth is chosen according to the phase diagrams of the respective metal - chalcogen systems [11]. The feed of the respective chalcogenide, sometimes with a small excess of the chalcogen, is placed in the hot part of the reaction vessel and gradually dissolves in the salt melt, migrates to the cold end of the ampoule, where it forms the crystals. CsCl-KCl-NaCl mixture of eutectic composition is the best to be used as the growth medium. The crystal growth lasts about three weeks.

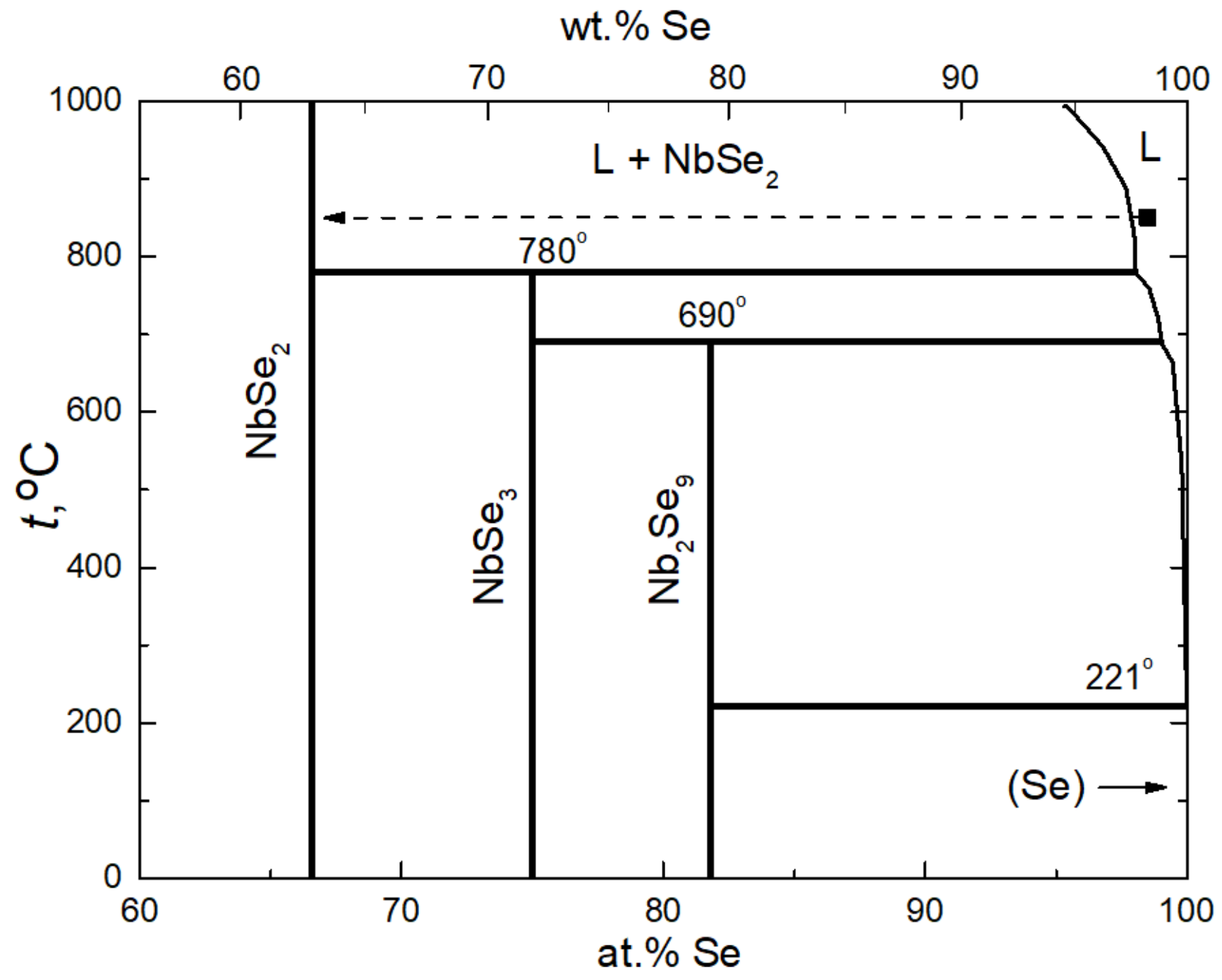

Figure 1. Partial phase diagram of the Nb-Se system [11]. The dashed arrow shows the gradual change in the system's composition in the hot end of the reaction vessel.

By now we obtained crystals of $TiS_2$, $TiSe_2$, $TiTe_2$, $ZrS_2$, $ZrSe_2$, $HfS_2$, $HfSe_2$, $VSe_2$, $MoSe_2$ and $WSe_2$ [16] using this technique, but the synthesis of dichalcogenides of niobium and noble metals was unsuccessful. Many transition metal dichalcogenides including niobium and noble metals are in equilibrium with molten chalcogen [11]. Often a chalcogen melt may contain significant amounts of the dissolved metal. For example, at 850°C, liquid selenium can contain up to four mass percent of niobium [11], Fig. 1. With such phase relationships, crystals can be easily obtained by solvent evaporation.

The application of evaporation can be demonstrated for the crystal growth of $RuS_2$ [17,18,19]. A boomerang-shaped silica glass tube (made by oxygen torch) was used as a reaction vessel, shown in Fig. 2a. Tellurium or selenium melt containing small amount of ruthenium or $RuS_2$, was placed in the left part of the vessel (growth zone) at 920°C, the solvent gradually evaporates and condenses at the right end of the ampoule at 900°C. Crystals of $RuS_2$ grew up to 3 mm in size in five days and small crystals of $RuTe_2$ or $RuSe_2$ were found in the growth zone. We continued the work on the growth of transition metal dichalcogenide crystals by evaporation method.

## CRYSTAL GROWTH

First of all, is it important to inform that the quartz ampoules, which contains an elementary chalcogenes and heated to high temperatures are extremely dangerous and unstable. The danger and unstable represented by hot fragments of both quartz glass and fumes of tellurium, selenium and sulfur. The ovens were located in isolated rooms. All manipulations with boomerangs were carried out with the protection of hands, face and respiratory organs. Tellurium (Alfa Aesar, 99.9999%), selenium (Alfa Aesar, 99.999%), sulfur (Labtex 99.9%) and metals with purity not less than 99.9% were used as the reagents. The feed for crystal growth was typically synthesized in advance in evacuated silica glass ampoules [20] from the elements taken in required proportions at the temperature approximately the same as planned for the crystal growth. Synthesized metal dichalcogenide powder was placed in a silica glass reaction vessel with the corresponding chalcogen. The reaction vessel was evacuated and sealed with an oxygen torch. Reaction vessels had the shape of a boomerang [17], the total length about 200 mm, inside diameter 8 mm and the thickness not less than 2.5 mm (Fig. 2a). Vessels were placed in a gradient furnace so that the metal-saturated chalcogen melt was at a higher temperature. The chalcogen gradually evaporated and condensed in the cold part of the vessel. Therefore, the percentage of the metal in the melt increased, which lead to the formation of the crystals of the phase most rich in chalcogen. Later the technique was slightly improved. At first the reaction vessel was located in the furnace so that the part with the chalcogen + metal melt was at a lower temperature (Fig. 2b). This allows to maximally dissolve the metal in the chalcogen melt without losing the chalcogen via evaporation. In a day or two after keeping the vessel in that position it is shifted to the working position (Fig. 2a) and the solvent evaporation process and the crystal growth begin. The similar temperature regime change can be also achieved by temperature inversion in a 2-zone gradient furnace. The crystal growth temperature is selected based on the two physical and chemical parameters:

- First of all, the crystal growth should proceed at temperatures sufficient for dissolving at least 1 wt.% of the metal in the chalcogen melt. Solubility is established by the position of the liquidus curve in the phase diagram.
- Second of all, the temperature should not be so high that the pressure would exceed the breaking limit for quartz reactors. Our experiments have shown that it is possible to confidently use ampoules with a non-aggressive substance inside with a pressure of 10 atm at 850°C. So, the low boiling sulfur can be used as a liquid solvent only below 600-650°C. Selenium vapor pressure above the liquid selenium at 685°C is one atmosphere and goes up to about 10 atmospheres at 850°C. Tellurium with boiling point of 980 °C can be used up to the highest temperatures.

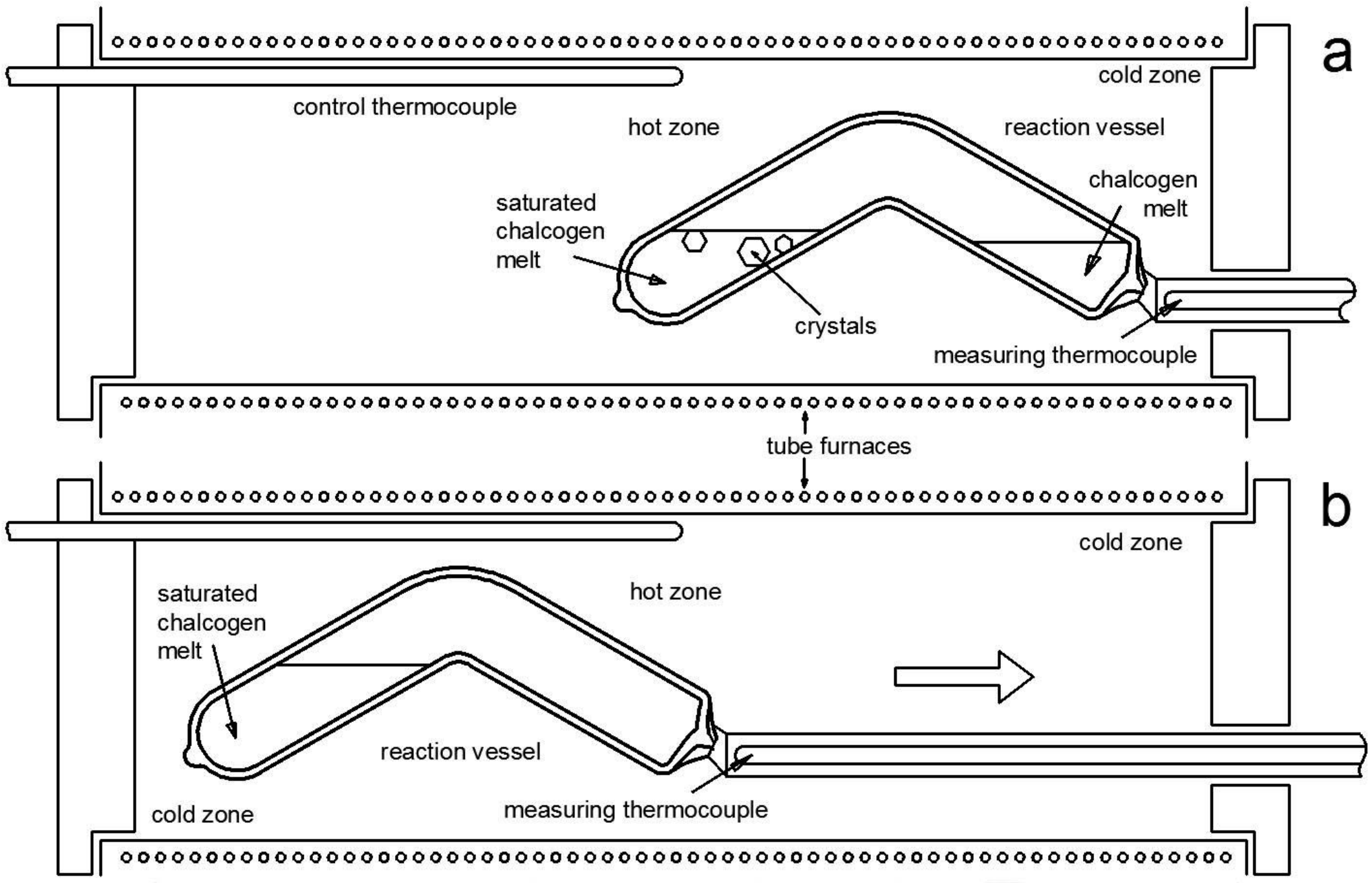

Figure 2. a) Silica glass reaction vessel in the shape of a boomerang for obtaining crystals of saturated chalcogenides by the chalcogen evaporation technique. The vessel is located in the furnace with a natural temperature gradient. The metal saturated chalcogen melt is on the left at a higher temperature. The chalcogen gradually evaporates and condenses in the cold part of the reaction vessel. b) A possible position of the reaction vessel before the beginning of the crystal growth. The chalcogen melt with the metal was located at a slightly lower temperature than the other end of the ampoule to make the metal dissolve completely. The arrow shows the shift of the reaction vessel for the chalcogen to start evaporating.

The temperature and the metal / chalcogen ratio were chosen according to the phase diagram. For example, $NbSe_2$ is in equilibrium with liquid selenium at temperatures above 780°C (see Fig. 1). At this temperature liquid selenium dissolves up to 1 wt% of niobium. At 850°C solubility of niobium in selenium is likely to reach 3 wt.%. Therefore 850°C is optimal for $NbSe_2$. Chemical composition of the crystals obtained was measured using Tescan Vega II XMU scanning electron microscope with INCA Energy 450 energy dispersive spectrometer in accelerating voltage of 20 kilovolts. Crystals glued to the conducting substrate as well as embedded into polished epoxy resin were studied. The crystals were ground and examined by x-ray powder diffraction technique on DRON-7 (Co$K_{\alpha}$-radiation, Fe-filter) or BRUKER (Cu$K_{\alpha 1}$-radiation, graphite monochromator) diffractometers. Crystals with apparent layered structure were checked for monocrystallinity on the BRUKER diffractometer.

All the experiments conducted to obtain crystals by the evaporation technique are summarized in table 1, see below. The table shows the composition of the feed, the approximate solubility of the metal in the melt according to the phase diagram, the temperatures of the hot and the cold ends of the reaction vessel, the time of the synthesis and the size of the crystals.

## I. Growth of ditelluride crystals

Amongst the chalcogens, tellurium is the most convenient and safe growth medium for obtaining dichalcogenide crystals because of its low vapor pressure and high solubility of metals in it. Thus, $PtTe_2$ crystals of millimeter size were obtained from the tellurium melt containing 5 wt.% of platinum in 700°C→560°C temperature regime (Fig. 3a). In similar temperature conditions, instead of faceted $PdTe_2$ crystals, a shapeless sintered mass was obtained. Melt evaporation in 655°C→543°C temperature regime allowed to obtain flat crystals of irregular shape with the size of up to 2 mm (Fig. 3b).

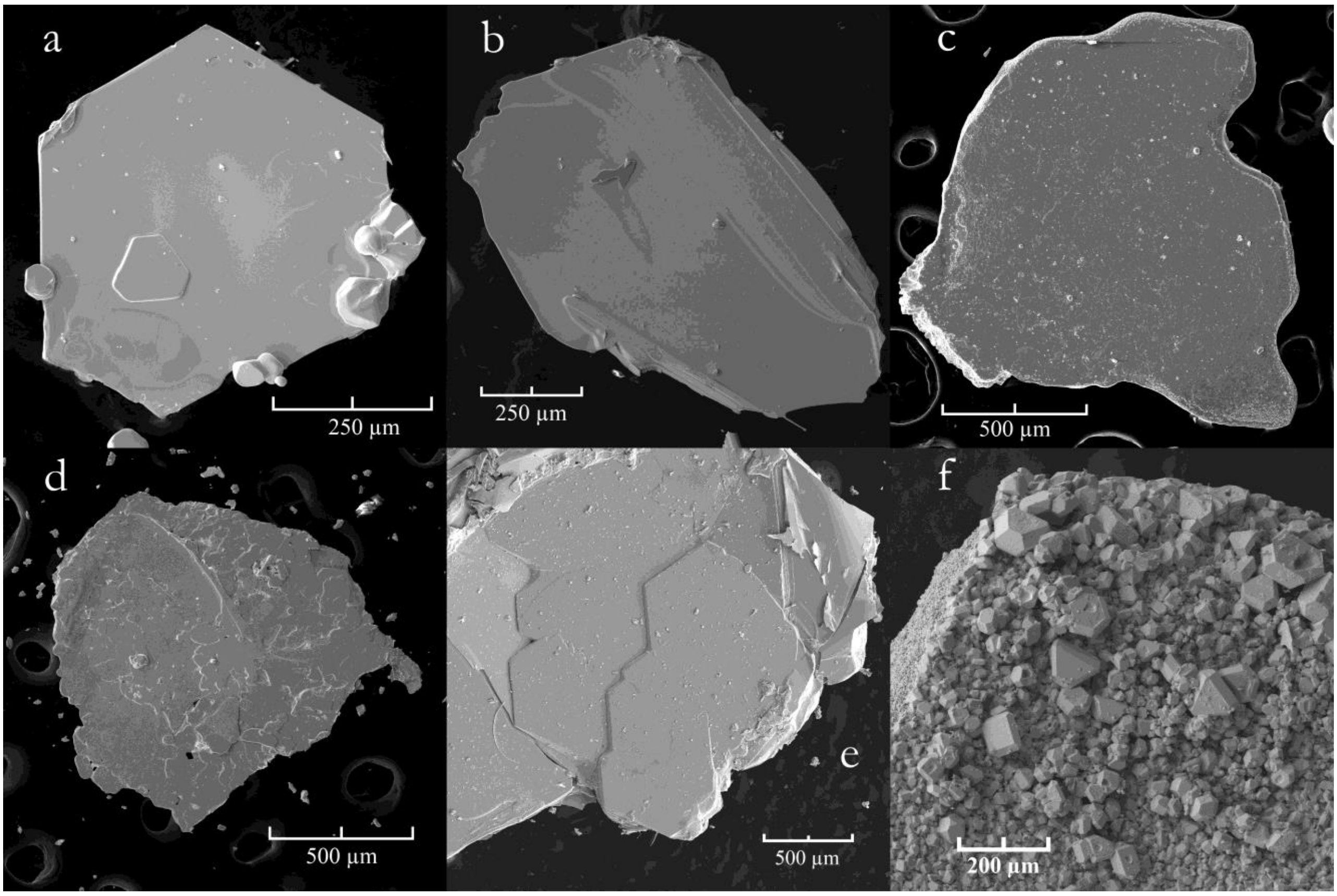

Figure 3. Electron microscope image of a crystal (or crystals) of $PtTe_2$ (a), $PdTe_2$ (b), $NiTe_2$ (c), $TaTe_2$ (d), $TiTe_2$ (e), $RuTe_2$ (f).

On tellurium evaporation from the melt with nickel at 720°C, with tantalum at 827°C, with titanium at 800-850°C, the substance with the apparent signs of recrystallisation was obtained, flat in shape, but with a slightly distinguishable crystal habit (Fig 3c, Fig 3d and Fig 3e). Probably the change in temperature regime, composition of the melt and other physical and chemical properties will allow to obtain crystals with a better habit. Low solubility of ruthenium in tellurium at 850°C did not allow to obtain $RuTe_2$ crystals greater than 100 microns in size (Fig. 3f). It is likely that the temperature increase or increase in total amount of the melt allows to obtain crystals larger in size [17-19]. Similarly, low rhenium solubility did not allow obtaining crystals in a similar temperature regime.

## II. Growth of diselenide crystals

Tablet-shaped $NbSe_2$ crystals were obtained by selenium evaporation from 850°C to 720°C. The sizes of the crystals were from several tenths of a millimeter up to one mm (Fig. 4a). Changes in the temperature of the cold end, boomerang shape and niobium concentration in the melt did not lead to noticeable change in size or shape of $NbSe_2$ crystals. In contrast, the shift of the reaction vessel (Fig. 2b), that leads to complete dissolution of niobium in selenium, allowed to increase the proportion of millimeter-size crystals.

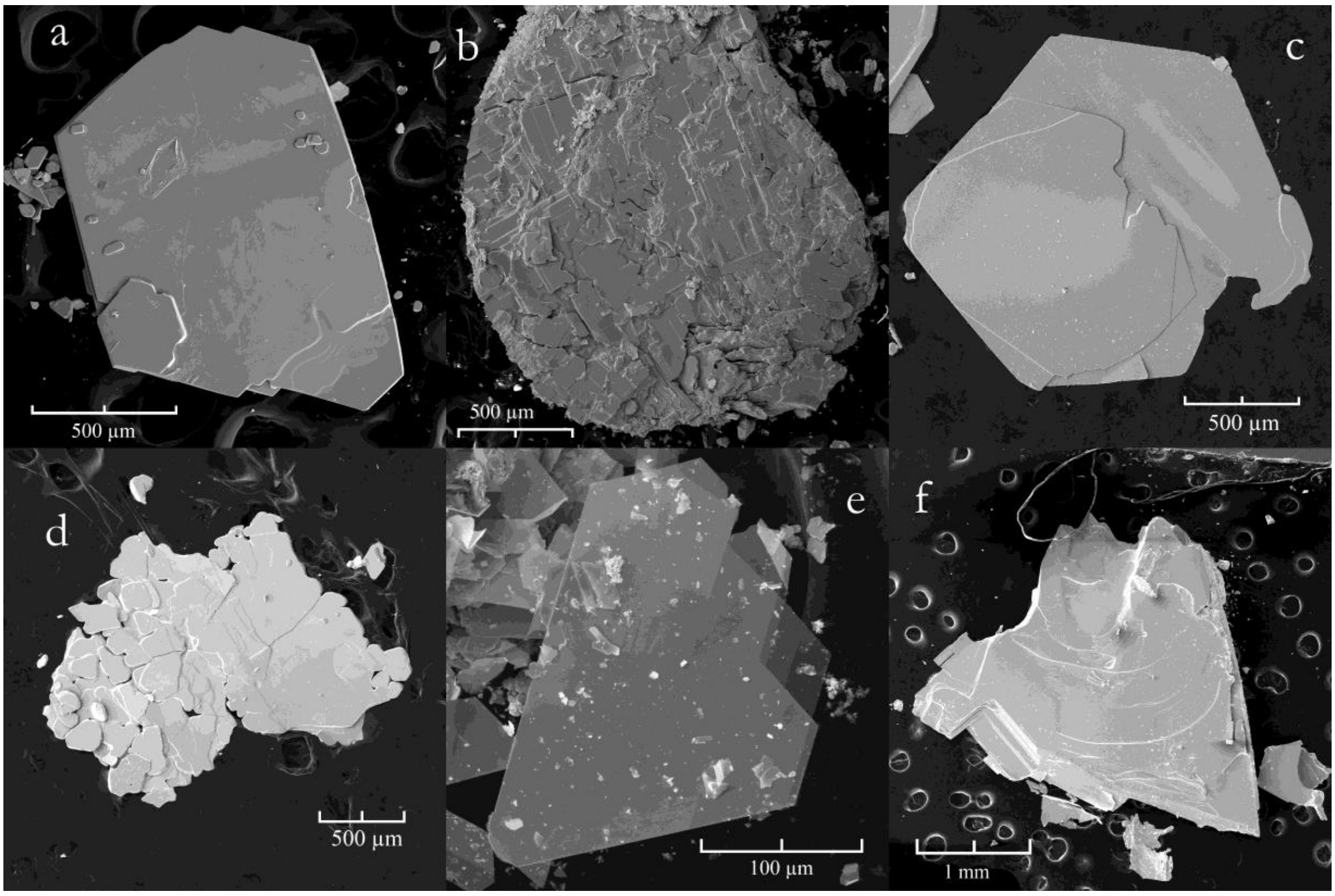

Figure 4. Electron microscope image of a crystal (or crystals) of $NbSe_2$ (a), $PdSe_2$ (b), $TiSe_2$ (c), $VSe_2$ (d), $ReSe_2$ (e), $Tb_2Te_5$ (f).

Occasionally, $NbO_3$ whiskers of length up to 15 mm were also found in reaction vessels near the condensed selenium. It is likely that oxygen got into the system in the form of selenium oxide or during silica glass recrystallization. The shift of the reaction vessel leading to the hot and the cold zones changing places allowed to increase the output in other experiments as well. Selenium evaporation from the melt with 2-2.3 % wt. Pd in regimes with the temperature of the hot end 650-730°C and of the cold end about 100°C lower has led to the formation of a round boule of palladium diselenide whose fracture shows the formation of an aggregate of oriented flat tetragonal crystals several hundred microns in size (Fig. 4b). Millimeter-size $TiSe_2$ crystals (Fig. 4c) were obtained by evaporation starting at 827°C. In similar conditions a sintered mass of oriented $VSe_2$ crystals (Fig. 4d) of up to 0.1 mm in size with a poorly formed crystal habit was obtained. Despite noticeable solubility of rhenium in selenium at 850°C the technique did not allow obtaining $ReSe_2$ crystals greater than two hundred microns in size (Fig. 4e). It is likely that the increase in temperature or in the total amount of the melt allows growing crystals of a greater size.

$PtSe_2$, $MoSe_2$ and $WSe_2$ crystals were not obtained at the temperature of the hot end of 827-850°C due to the low solubility of platinum, molybdenum and wolfram in liquid selenium. Practically the complete absence of any traces of $PtSe_2$ powder dissolution in liquid selenium contradicts the schematic phase diagram of the Pt-Se system [22]. In addition, $Tb_2Te_5$ crystals were obtained by evaporation technique in the temperature regime of 790°C→660°C (Fig. 4f). In general, liquidus curve shapes in REE-chalcogen systems should allow obtaining $REE_2Ch_5$ or $REECh_3$ (Ch – Se, Te) crystals.

## III. Growth of ternary compound crystals

Besides obtaining simple dichalcogenides there is a possibility to obtain crystals doped with other cations. For example, $NiTe_2$ crystals doped with palladium, $VSe_2$, crystals doped with nickel and $NbSe_2$ crystals doped with iron (Table 1). $NbSe_2$ crystals can be obtained as a single phase as well as in equilibrium with iron selenide (Fig. 5a). Replacing a part of the chalcogen with another anion is also possible. For example $Pt(Te,Bi)_2$ crystals were obtained with about 40% of tellurium replaced by bismuth. It can be assumed that only tellurium evaporates in this system while bismuth remains in the melt due to its low partial vapor pressure.

On the contrary, when obtaining sulfur containing crystals, during evaporation of saturated selenium or tellurium melt several opposite factors should be considered. Firstly, metals typically have greater affinity to sulfur which leads to a stable "metal sulfide – liquid Se (or Te)" association. Secondly, according to the first Konovalov's law, the gas phase should be rich and the liquid phase poor in sulfur. Thirdly, sulfur can act as a transport agent and carry metals to the cold end or to the middle of the reaction vessel. Complicated chemistry and a large number of factors require further studying of crystal growth in systems containing sulfur. For example, during evaporation of the Nb-Se-S system $NbS_2$ crystals were obtained with 25 to 3% of sulfur replaced by selenium. Crystals with the smallest selenium content (Fig. 5b) were obtained in the system with a large amount of sulfur and therefore with the low temperature of the cold zone (540°C) in order to reduce the total pressure (Table 1).

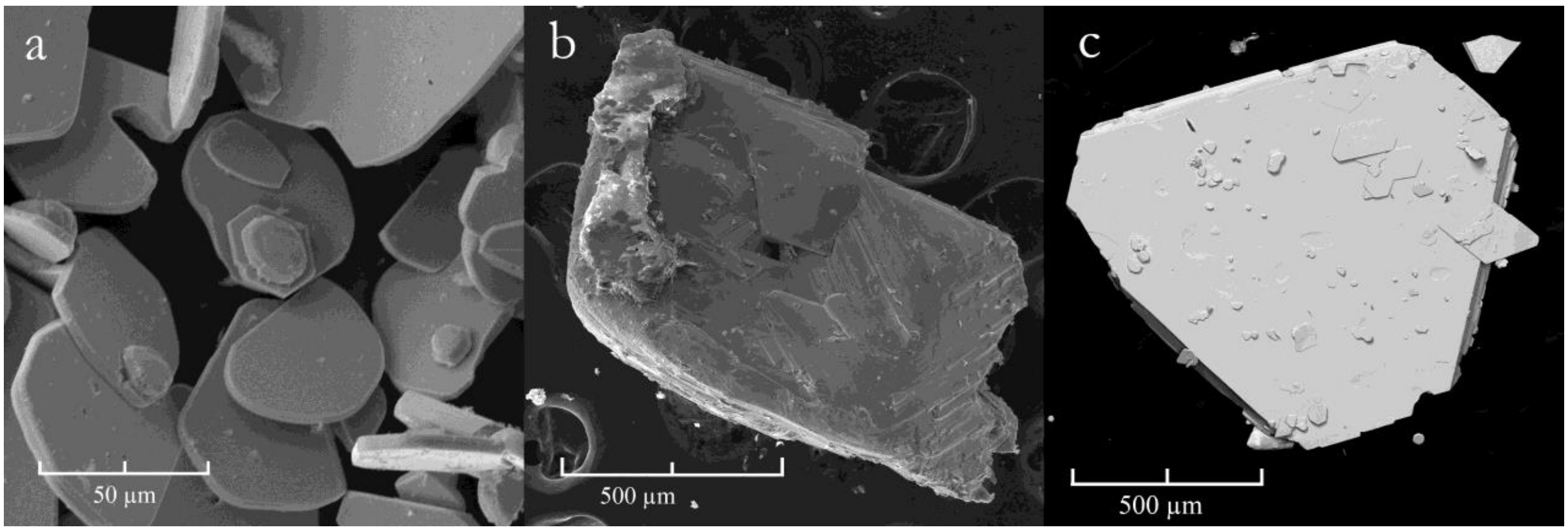

Figure 5. Electron microscope image of $NbSe_2$ crystals in equilibrium with $Fe_{1-\delta}Se$ (in the middle) (a), $Nb(S,Se)_2$ (b), $PtSe_2$ (c)

Besides the third component might not incorporate into the growing crystal but increase the metal solubility in the evaporating melt. This way millimeter $PtSe_2$ crystals were obtained during $Hg_{0.5}Se_{0.5}$ melt evaporation in the regime 850°C→720°C (Fig. 5c). It is likely that molten mercury selenide can be used as a solvent to obtain selenides of other metals. The lowest temperature at which the melt can be used is HgSe congruent melting point - 799°C, the highest relatively safe temperature at which the total pressure of mercury and selenium will not exceed ten atmospheres is ~ 870°C [23].

In the images (3d, 4b, 4d) it can be seen that despite the small size of the crystals they have the form of associations oriented along the c axis, which allows measuring some anisotropic properties.-Thus, evaporation technique is a decent alternative to other flux synthesis techniques. For example, when alkali metal halogenides are used as a solvent some transition metals might not be transported [16]. Furthermore, the alkali metal can incorporate into the structure of the growing crystal. Besides, the technique has other advantages: Using the chalcogen melt in gradient conditions, when the powder charge dissolves in liquid chalcogen in the cold end of the reaction vessel, migrates and crystallizes in the cold end [24], is also less convenient and less safe than evaporation: the average temperature of the reaction vessel should be higher. The quality of the reaction products can be evaluated without opening the reaction vessel and if necessary the growth experiment can be redone. Crystals obtained do not require washing from the remaining solvent. The remaining chalcogen and the reaction vessel can be used repeatedly.

## CHARACTERIZATION OF SELECTED CRYSTALS

To assess the quality of some grown crystals, we have characterized various physical properties of selected crystals. Six terminals were used to determine each principal component of resistivity in the *ab* plane, $\varrho_{ab}$, and along the *c* axis, $\varrho_c$ at ambient pressure. Low-temperature transport and specific heat (down to 400 mK) were measured with a Physical Property Measurement System (PPMS, Quantum Design).

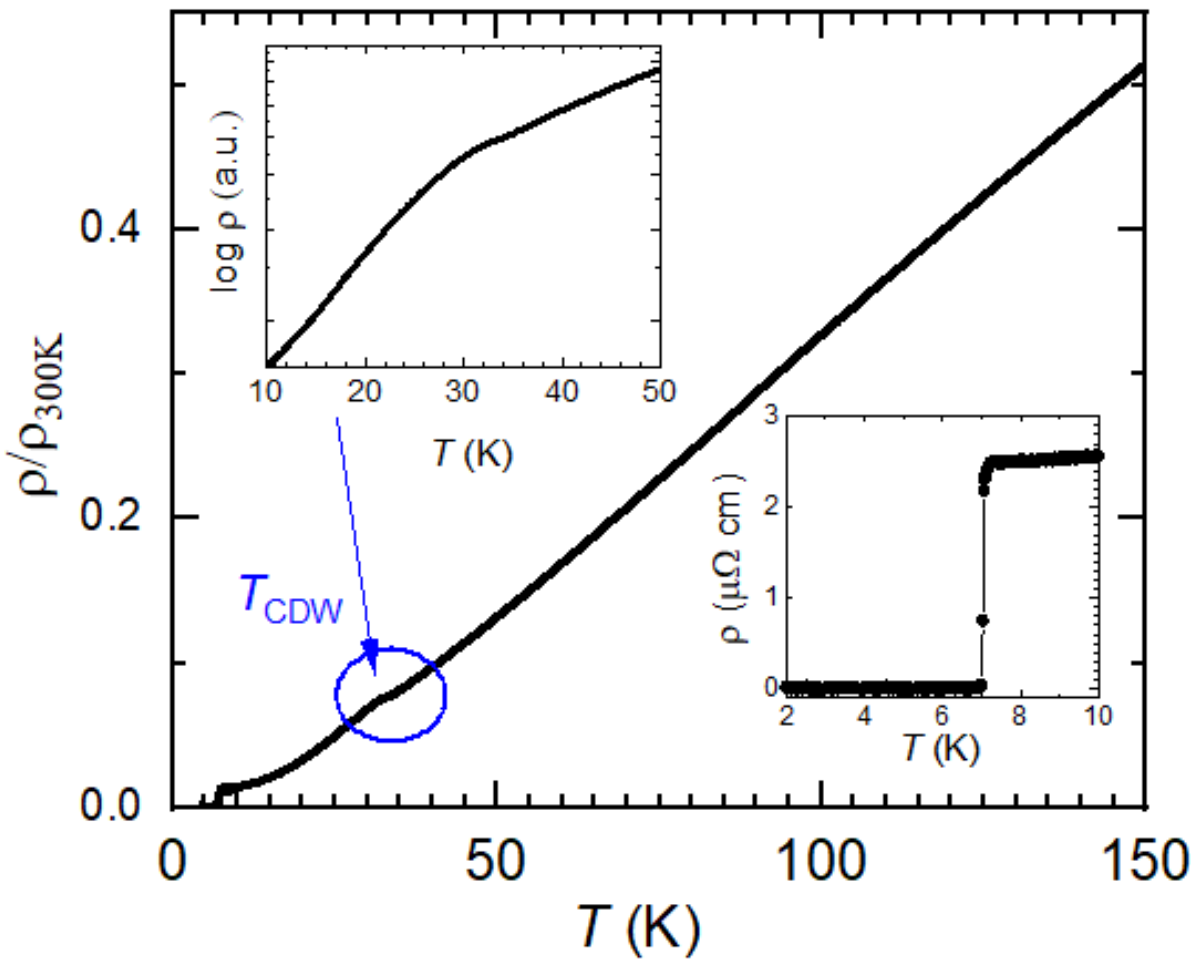

Figure 6. Temperature dependence of the in-plane resistivity measurements for 2H-$NbSe_2$ single crystals during heating up to room temperature (300 K). The inset is the close-up of the low-temperature region, highlighting the CDW formation in $NbSe_2$.

Figure 6 illustrates the temperature dependence of the in-plane resistivity measurements of $NbSe_2$ single crystals during warming from 0.4 K to room temperature (300 K). The normal state exhibits simple metallic behavior during cooling down from room-*T*, followed by a sharp superconducting transition at $T_c$, which agrees with magnetization and specific heat data (described later). The resistivity curve of 2H-$NbSe_2$ presents an anomaly near 33 K, which is attributed to the formation of the CDW. The inset of Fig. 6 is the close-up of the low-temperature region, highlighting the CDW formation. One can notice that the normal state exhibits a simple metallic behavior upon cooling down to room temperature, followed by a sharp SC transition at Tc 7.2 K (90% of the normal state). The value of residual resistivity ratio (RRR) is 78. Both the large RRR and narrow superconducting transition confirm the good quality of the samples investigated here. Low-temperature specific heat, $C_P$, being equal to the temperature derivative of the entropy, probes the gap structure of bulk superconductors. The inset of (Fig.7) illustrates the temperature dependence of the specific heat of $NbSe_2$ measured at different applied magnetic fields parallel to the *c*-axis. In zero-field specific-heat measurements, a very sharp anomaly is clearly seen, which is attributed to the superconducting transition at $T_c$. This specific-heat jump is systematically shifted to lower temperatures upon applying DC magnetic fields of up to 6 T. The main panel of Fig. 7 presents a bump near 33 K for $NbSe_2$, which is due to the formation of the CDW, followed by a sharp SC transition at $T_c = 7.2$ K, agreeing with magnetization and specific heat data previously reported.

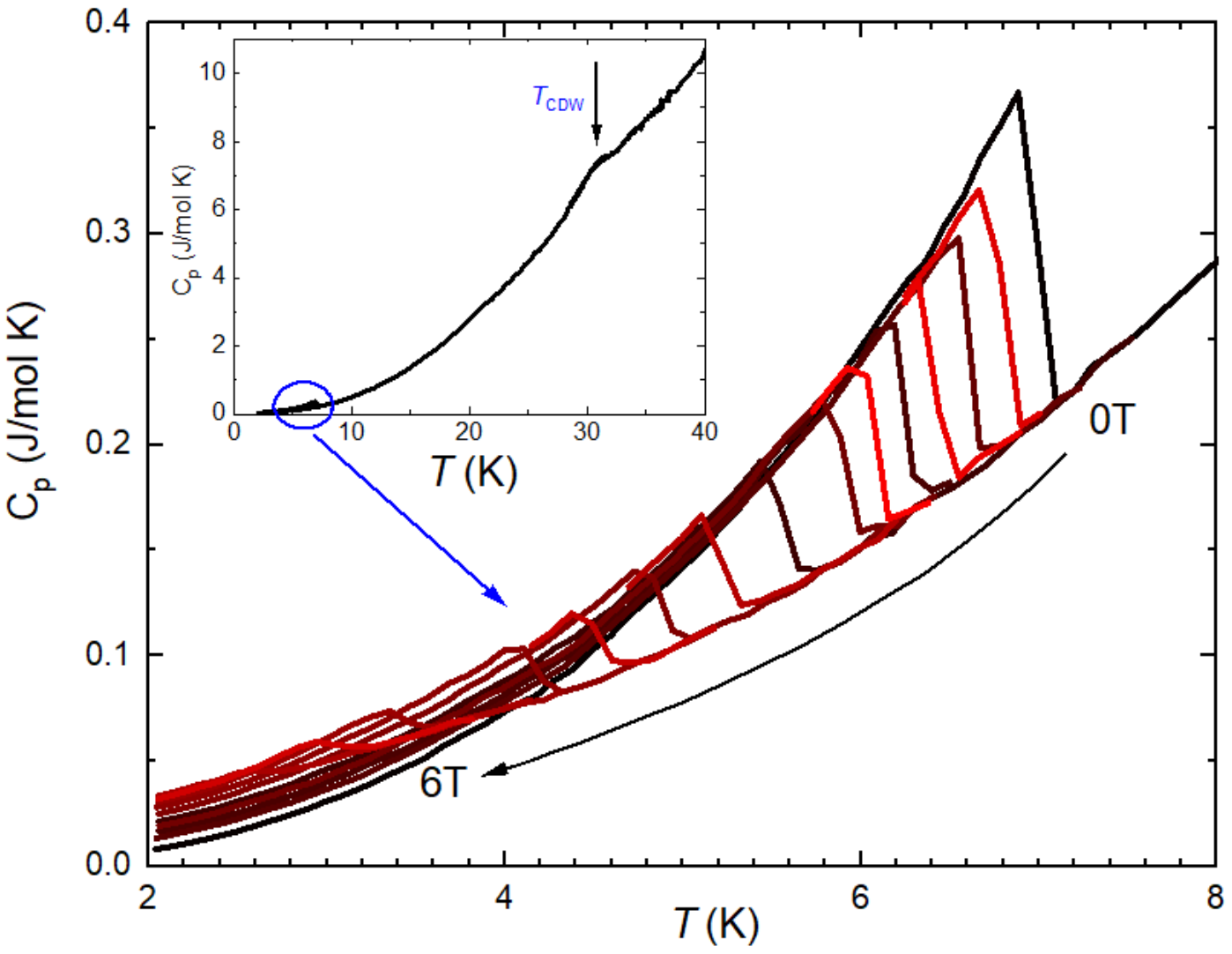

Figure 7. The main panel shows the close-up of the low-temperature superconducting region in 2H- $NbSe_2$ of the specific heat measured at different applied magnetic fields parallel to the *c*-axis, with increments of 0.5 T. The inset presents the temperature dependence of specific heat measured at ambient pressure. The zero field measurements illustrates the CDW formation in $NbSe_2$.

The temperature dependence of the resistivity at zero field of pristine 1T-$TiSe_2$ is shown in the main panel of Fig. 8. Interestingly, one can see that the temperature dependence exhibits a maximum close to 162 K (see the lower inset of Fig.8), which has previously been suggested to arise from an initial decrease in the density of available carriers[25]. This is caused by the opening of a gap in the charge-ordered phase, which is overtaken at lower temperatures by both the decrease of scattering channels due to the developing order and an increase in density of states due to the downward shift of the conduction band minimum below $T_{CDW}$. However, in the CDW state, only a fraction of the states at Fermi energy, $E_F$, are gapped and the density of states (DOS) near $E_F$ directly correlates with the ordering strength. The position of the maximum thus does not coincide with any charge-ordering transition. However, this striking feature of the obtained CDW order will be discussed in a separate study. In order to investigate the effect of the magnetic field *H* on the CDW in $TiSe_2$, we measured the resistivity with the magnetic field applied along the *c*-axis, as shown in the upper inset of Fig. 8 (upper inset). However, no change in the CDW anomaly has been found. It suggests that the effect of the applied magnetic field on the CDW transition is negligible.

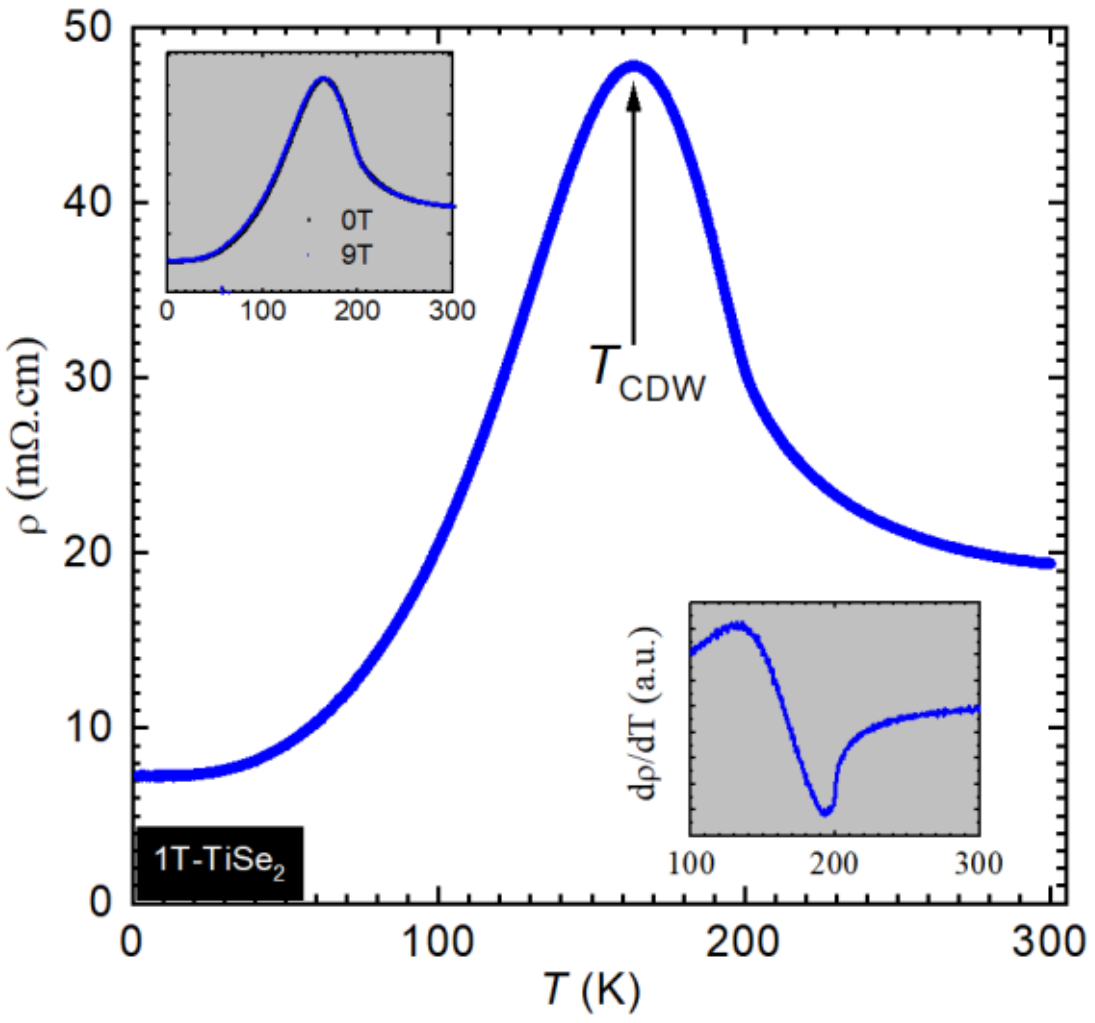

Figure 8. The temperature profiles of resistivity. The lower inset presents the derivative of the data in which the CDW transition occurs. The upper inset illustrate the effect of the magnetic field on the CDW transition.

The main panel of Fig. 9 illustrates the temperature dependence of the heat capacity measurements of $PtTe_2$ single crystals during heating in zero field. For $PtTe_2$, the Dulong-Petit value reaches $3RN$=74.9 J/mol K, with the number of atoms per formula unit, $R$ =3 and the gas constant $R$=8.314 J/mol K. The zero-field specific-heat data can be plotted as Cp/T versus $T^2$ following $C_{el}$ + $C_{ph}$ = $\gamma_n T + \beta T^3$, with $\gamma_n$ and $\beta$ as the electronic and lattice coefficients (see inset of Fig. 9). The obtained value of $\gamma_n$ for the investigated sample is $\gamma_n$ = 2.7(2) mJ/mol $K^2$. The phononic coefficient $\beta$ is found to be 0.608 mJ/mol $K^4$. Using the relation $\theta_D = (12\pi^4 RN/5\beta)^{1/3}$, where R is the molar gas constant and N = 3 is the number of atoms per formula unit, the Debye temperature $\theta_D$ = 134 (5) K is obtained. Remarkably, the fact that the low-temperature specific-heat data for $PtTe_2$ exhibit a linear behavior at low temperatures without any upturn indicates the absence of Schottky-like contributions in the investigated crystal.

The electronic structures of two-dimensional TMDs show remarkably exotic ground states that are only beginning to be discovered. This is particularly interesting in compounds with nontrivial topological properties that explicitly influence linear electronic dispersions in the band structure of the solid. $PtTe_2$ is among the rare compounds to show type-II Dirac fermions, whereby charge carriers with linear bands and massless properties emerge at symmetry protected points or along nodal-lines in the momentum space. Angle-resolved photoemission spectroscopy and electron diffraction measurements are used to directly access the electronic and crystal structures of $PtTe_2$. The hexagonal structure of $PtTe_2$ consists of three covalently bonded Te−Pt−Te atomic planes in the prototypical $CdI_2$ structure as shown in Fig. 10(a) [26]. Low-energy electron diffraction (LEED) in confirms an excellent (1 x 1) hexagonal pattern without any surface contamination. This ensures a high-quality single crystal with a global surface quality that was cleaved *in-situ* prior to ARPES measurements.

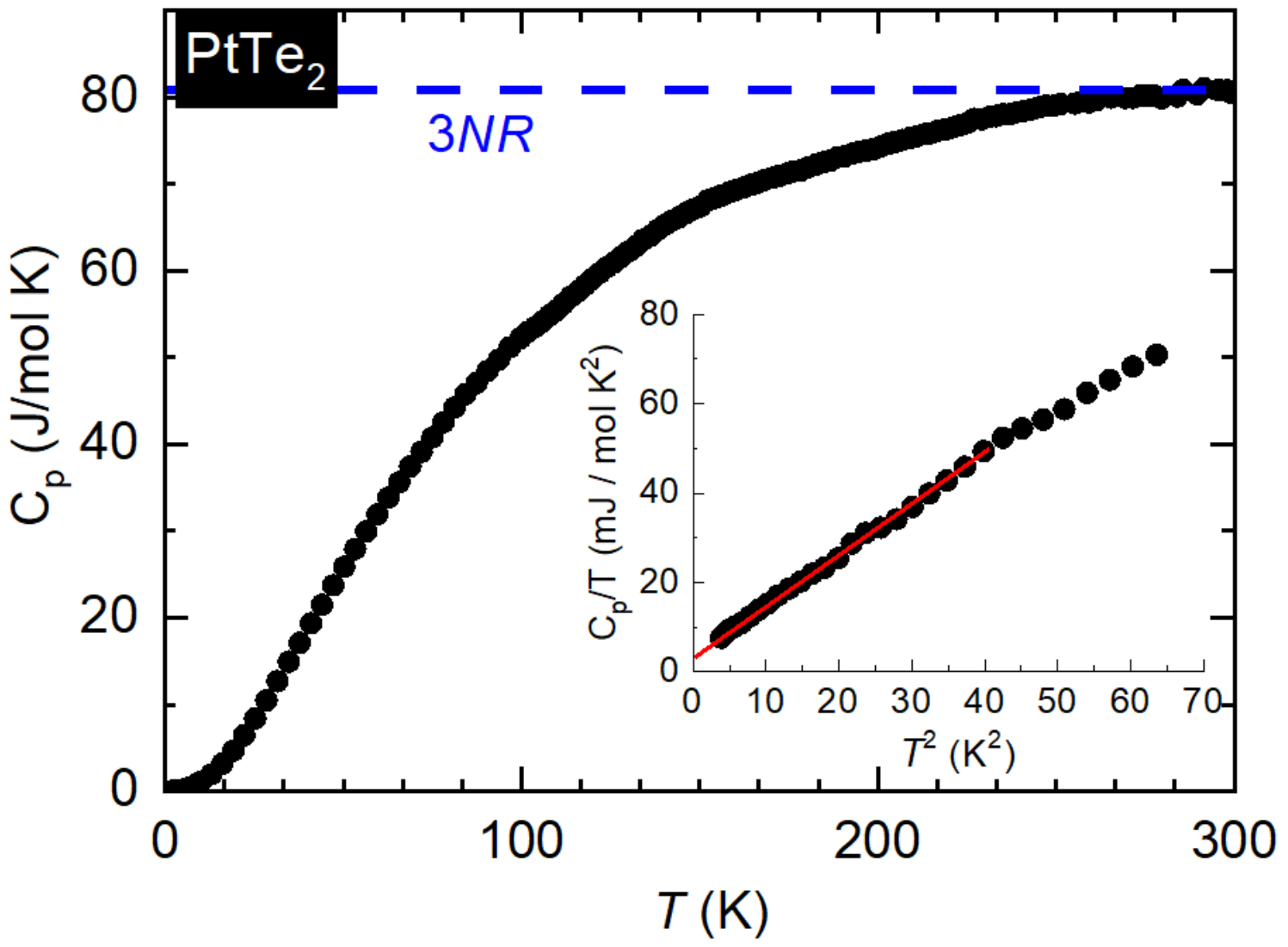

Figure 9. the temperature dependence of the heat capacity measurements of $PtTe_2$ single crystals during heating in zero field up to room temperature with no other phase transition upon the whole temperature range. The inset shows the plot C/T versus $T^2$. The solid red line shows linear fits to $C/T = \gamma_n + \beta T^2$.

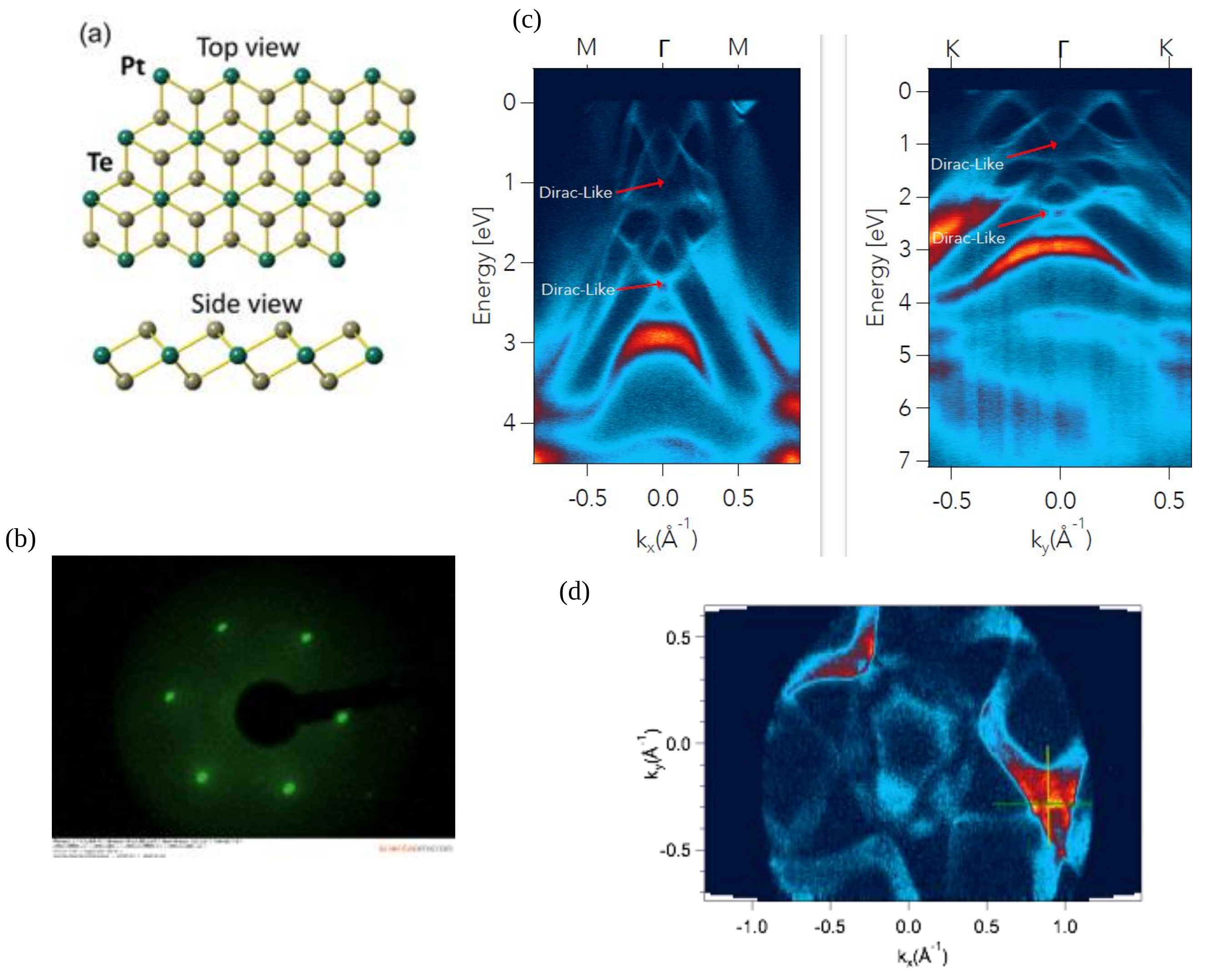

Figure 10: Angle-resolved photoemission spectroscopy and electron in $PtTe_2$. The hexagonal structure of $PtTe_2$ consists of three covalently bonded Te−Pt−Te atomic planes in the prototypical $CdI_2$ structure as shown in (a). Low-energy electron diffraction (LEED) in (b) confirms an excellent (1 x 1) hexagonal pattern without any surface contamination. The overview of $PtTe_2$ band structure measured by ARPES is shown in (c). An additional electron pocket is also seen in between the Γ-K direction in (d).

Table 1. Parameters of experiments on the dichalcogenides growth: particular composition of the feed, solubility approximation of the metal in melt according to the phase diagram, temperatures of both hot and cold ends of the reaction vessel, time of the synthesis and size of the crystals.

| crystals, crystal structure if not $CdI_2$ | melt composition, wt.% | Metal solubility in melt (wt.%) at T of evaporation[11] | T°C of hot end (of evaporation) | T°C of cold end (of condensation) | synthesis time, days | crystal size | Fig. № |
|---|---|---|---|---|---|---|---|
| $PtTe_2$ | 5% Pt | ~5% | 700 | 590 | 5 | 1-2 mm | 3a |
| $PdTe_2$ | 9.3% Pd | ~12% | 700 | 590 | 5 | mm agglomerates | |
| $PdTe_2$ | 9.4% Pd | ~8% | 655 | 543 | 7 | 2 mm | 3b |
| $NiTe_2$ | 7.5% Ni | ~8% | 720 | 600 | 5 | mm<br>agglomerates | 3c |
| $TaTe_2$ | 5% Ta | n/d | 827 | 720 | 6 | mm<br>agglomerates | 3d |
| $TiTe_2$ | 5% Ti | ~4% [21]<br>~3% | 850<br>800 | 750<br>700 | 5 | 1 mm | 3e |
| $RuTe_2$<br>($FeS_2$ str.) | 4% Ru | < 1% | 850 | 720 | 5 | 10-15 μm | 3f |
| $ReTe_2$ | 4.5 | ~5% | 850 | 750 | 5 | no trace | |
| $NbSe_2$ | 4% Nb | 3% Nb | 850 | 720 | 5 | 1 mm | 4a |
| $PdSe_2$<br>($FeS_2$ str.) | 2% Pd | ~4%<br>~5% | 700<br>730 | 590<br>620 | 6 | 50 μm | <br>4b |
| $TiSe_2$ | 4.3% Ti | n/d | 827 | | 6 | 1 mm | 4c |
| $VSe_2$ | 18% V | ~15% | 827 | | 7 | 0.1 mm | 4d |
| $ReSe_2$ | 5% Re | n/d | 850 | 750 | 7 | 0.2 mm | 4e |
| $PtSe_2$ | 5% Pt | [22] | 850 | 720 | 5 | no trace | |
| $MoSe_2$<br>$P6_3/mmc$ | 5% Mo | n/d | 850 | 750 | 7 | no trace | |
| $WSe_2$<br>($MoS_2$ str.) | 7% W | n/d | 827 | | 8 | no trace | |
| $Tb_2Te_5$<br>($Nb_2Te_5$ str.) | 3.7% Tb | n/d | 790 | 660 | 7 | ~ 1 mm | 4f |
| $Pd_{0.15}Ni_{0.85}Te_2$ | 2%Pd, 6%Ni | n/d | 700 | 600 | 5 | ~ 1 mm | |
| $V_{0.88}Ni_{0.12}Se_2$ | 15% V<br>1.7% Ni | n/d | 750 | 640 | 5 | 100 μm | |
| $NbSe_2$:Fe | 3%Nb 0.01%$^{57}$Fe | n/d | 850 | 730 | 5 | 100 μm | |
| $Nb_{0.95}Fe_{0.05}Se_2$ | 4%Nb 0.1%Fe | n/d | 850 | 720 | 5 | 100 μm | |
| $Nb_{0.86}Fe_{0.14}Se_2$ | 4%Nb 0.5%Fe | n/d | 850 | 730 | 5 | 100 μm $NbSe_2$<br>20 μm FeSe | 5a |
| $Pt(Te_{0.6}Bi_{0.4})_2$ | 5%Pt, 4%Bi | n/d | 700 | 610 | 6 | 1-3 mm | |
| $Nb(S_{0.75}Se_{0.25})_2$ | 4%Nb, 3%S | n/d | 850 | 720 | 6 | 500 μm | |
| $Nb(S_{0.88}Se_{0.12})_2$ | 4%Nb, 3%S | n/d | 850 | 720 | 5 | 500 μm | |
| $Nb(S_{0.97}Se_{0.03})_2$<br>$MoS_2$ str.? | 18% Nb<br>31% Se | ~9% | 850 | 540 | 30 | 1 mm | 5b |
| $PtSe_2$ | 3% Pt<br>44 % Hg | n/d | 850 | 720 | 6 | ~ 1 mm | 5c |

## CONCLUSIONS

Here we report a method for growing high quality crystals of transition metal diselenide and ditelluride compounds via chalcogen solvent evaporation, which is an alternative method for obtaining crystals with high chalcogen content. Several compounds which we were unable to synthesize from alkali metal melts can be obtained using the reported method [16]. ARPES and LEED characterization of $PtTe_2$ crystals grown using the reported method indicate that the crystal quality is comparable to crystals grown using other state-of-the-art methods. Very recently, report on high quality complementary measurements for 2H-$NbSe_2$ and 2H-$NbSe_2$ show that the estimated values of the penetration depth at $T$ = 0K depart from a Uemura-style relationship between $Tc$ with $\lambda^{-2}_{ab}$ ($T$), the in-plane superconducting penetration depth [30].

## ACKNOWLEDGMENTS

The crystal growth experiments were supported by the Russian Science Foundation (Project No.19-12-00414). The work of DACh is supported by the program 211 of the Russian Federation Government, agreement No. 02.A03.21.0006, by the Russian Government Program of Competitive Growth of Kazan Federal University. The work of ANV was supported by the Ministry of Education and Science of the Russian Federation in the framework of Increase Competitiveness Program of NUST "MISiS" grant K2-2017-084, by acts 211 of the Government of Russian Federation, Contracts No. 02.A03.21.0004, 02.A03.21.0006 and 02.A03.21.0011. The work in Germany was supported by the program MO30214/2. MAH acknowledges support from the VR starting grant 2018-05339. We acknowledge MAX IV Laboratory for time on Beamline Bloch under Proposal 20190335. Research conducted at MAX IV, a Swedish national user facility, is supported by the Swedish Research council under contract 2018-07152, the Swedish Governmental Agency for Innovation Systems under contract 2018-04969, and Formas under contract 2019-02496. We acknowledge ARPES experiment support from Craig Polley (MAX IV), Maciej Dendzik (KTH), Antonija Grubisic-Cabo (KTH) and Oscar Tjernberg (KTH). HR, DP and GJM acknowledge the Swedish Research Council (2018-06465, 2018-04330) and the Swedish Energy Agency (P43549-1) for financial support.